\documentclass[11pt]{article}
\pdfoutput=1
\usepackage[english]{babel}
\usepackage{graphicx}
\usepackage{amsmath}
\usepackage{amssymb}
\usepackage{wrapfig}
\usepackage{subfigure}
\usepackage{caption2}
\usepackage{cite}
\begin{document}
\renewcommand{\baselinestretch}{1.346} %for preprints only

\setcounter{equation}{0} \setcounter{figure}{0}
\setcounter{footnote}{0}
\title{\bf Response, relaxation and transport \\
in unconventional superconductors}
\author{Dietrich Einzel and Ludwig Klam \\
{\it Walther--Meissner--Institut f\"ur Tieftemperaturforschung, D--85748 Garching, FRG}}

\date{\today}
\maketitle
% Formula macros #####################################################
\def\bdm{\begin{displaymath}} \def\edm{\end{displaymath}}
\def\nn{\nonumber} \def\bc{\begin{center}} \def\ec{\end{center}}
\def\be{\begin{equation}} \def\ee{\end{equation}}
\def\tcb{\textcolor{blue}} \def\tcbl{\textcolor{black}}
\def\tcg{\textcolor{green}} \def\tcr{\textcolor{red}}
\def\tcgr{\textcolor{grey}} \def\va{{\bf a}} \def\vA{{\bf A}} \def\vb{{\bf b}}
\def\vB{{\bf B}} \def\vb{{\bf b}} \def\vc{{\bf c}} \def\vC{{\bf C}}
\def\vd{{\bf d}} \def\hvd{\hat\vd} \def\vD{{\bf D}} \def\ve{{\bf e}}
\def\hve{\hat\ve} \def\vE{{\bf E}} \def\vf{{\bf f}} \def\vF{{\bf F}}
\def\vg{{\bf g}} \def\vG{{\bf G}} \def\vh{{\bf h}} \def\vH{{\bf H}}
\def\vi{{\bf i}} \def\vI{{\bf I}} \def\vj{{\bf j}} \def\vJ{{\bf J}}
\def\vk{{\bf k}} \def\hvk{\hat\vk} \def\vK{{\bf K}} \def\vl{{\bf l}}
\def\vL{{\bf L}} \def\vLambda{{\bf\Lambda}} \def\vm{{\bf m}} \def\vM{{\bf M}}
\def\vn{{\bf n}} \def\hvn{\hat\vn} \def\vN{{\bf N}} \def\vone{{\bf 1}}
\def\vp{{\bf p}} \def\hvp{\hat\vp} \def\vP{{\bf P}} \def\vq{{\bf q}}
\def\vQ{{\bf Q}} \def\vr{{\bf r}} \def\vR{{\bf R}} \def\vs{{\bf s}}
\def\vS{{\bf S}} \def\vt{{\bf t}} \def\vT{{\bf T}} \def\vu{{\bf u}}
\def\vU{{\bf U}} \def\vv{{\bf v}} \def\vV{{\bf V}} \def\vw{{\bf w}}
\def\vW{{\bf W}} \def\vx{{\bf x}} \def\vX{{\bf X}} \def\vy{{\bf y}}
\def\vY{{\bf Y}} \def\vz{{\bf z}} \def\v0{{\bf 0}} \def\hvz{\hat\vz}
\def\vZ{{\bf Z}} \def\vtau{{\bf \tau}} \def\e{{\rm e}} \def\kB{k_{\rm B}}
\def\kF{k_{\rm F}} \def\EF{E_{\rm F}} \def\NF{N_{\rm F}} \def\pF{p_{\rm F}}
\def\Tc{T_{\rm c}} \def\vvF{v_{\rm F}} \def\vna{{\bf\nabla}} \def\vPi{{\bf\Pi}}
\def\vgamma{{\mbox{\boldmath$\gamma$}}} \def\vsigma{{\mbox{\boldmath$\sigma$}}}
\def\vPi{{\mbox{\boldmath$\Pi$}}} \def\vzeta{{\mbox{\boldmath$\zeta$}}}
\def\vchi{{\mbox{\boldmath$\chi$}}}

\begin{abstract}
We investigate the collision--limited electronic Raman
response and the attenuation of ultrasound in spin--singlet
$d$--wave superconductors at low temperatures. The dominating
elastic collisions are treated within a t--matrix approximation,
which combines the description of weak (Born) and strong (unitary)
impurity scattering. In the long wavelength limit a two--fluid
description of both response and transport emerges. Collisions are
here seen to exclusively dominate the relaxational dynamics of the
(Bogoliubov) quasiparticle system and the analysis allows for a
clear connection of response and transport phenomena. When applied
to quasi--2--$d$ superconductors like the cuprates, it turns out
that the transport parameter associated with the Raman scattering
intensity for B$_{1g}$ and B$_{2g}$ photon polarization is closely
related to the corresponding components of the shear viscosity
tensor, which dominates the attenuation of ultrasound. At low
temperatures we present analytic solutions of the transport
equations, resulting in a non--power--law behavior of
the transport parameters on temperature. 
\end{abstract}

\section{Introduction}
During the last few decades a large
variety of so--called {\it unconventional} superconductors have been
discovered, among these the superfluid phases of $^3$He \cite{OLR,LEGGETT75,WHEATLEY75,VANDW}, the heavy Fermion
systems\cite{STEGLICHETAL,OTTETAL1,STEWETAL}, the
cuprates\cite{BANDM,TANDK00} and the Ruddlesden--Popper system
Sr$_2$RuO$_4$\cite{MAENOETAL}. The unconventional pairing
correlations in these systems manifest themselves on the one hand in
an anisotropy of the pair potential or energy gap of less symmetry
than the underlying band structure or in the occurrence of
additional spontaneously broken symmetries besides the $U(1)$ gauge
symmetry. On the other hand their existence can be detected from the
sensitivity of thermodynamic quantities like the transition
temperature or the equilibrium energy gap to even small amounts of
non--magnetic impurities. \\

In an earlier publication \cite{DE03}, a simple two--fluid
description for these unconventional superconductors was formulated,
which emerges from the BCS theory in the long wavelength ($\vq\to
0$) and stationary ($\omega\to 0$) limit, sometimes also referred to
as the {\it local equilibrium}. As a result, even analytical results
were obtained for the temperature dependence of the local reactive
response functions of the normal component, the Bogoliubov
quasiparticles (specific heat capacity, spin susceptibility) and the
condensate (superfluid density, magnetic penetration depth).
Clearly, a comprehensive two--fluid description of superconductors
should describe the more general situation beyond local equilibrium
and should therefore contain the dissipative response of both the
quasiparticle system and the condensate. A first step in this
direction was the derivation of a general form of a certain class of
quasiparticle transport parameters for unconventional
superconductors in reference \cite{EANDP05}. The results for the
impurity--limited transport of momentum (shear viscosity) and energy
(diffusive thermal conductivity) were, however, discussed
exclusively for the case of superfluid $^3$He--B with silica aerogel
forming the impurity system. Moreover, the two aspects of response
and transport traditionally appear as fairly remote aspects of
the reactive and dissipative dynamics of a superconductor. \\

Therefore, this paper is devoted to a unified description of
response and transport in superconductors. To be specific, we limit
our considerations to unconventional spin--singlet superconductors
with $d$--wave pairing correlations, in view of an applicability to
hole--doped cuprate superconductors. We would furthermore like to
concentrate on the electronic Raman \cite{DANDH07} and stress tensor
response. When treated in the long wavelength limit, the
corresponding response functions can be shown to be separable into
normal (Bogoliubov quasiparticles) and superconducting (pair
condensate) contributions, respectively, hence allowing for a
two--fluid description at {\it arbitrary} quasiclassical frequencies
in the homogeneous ($\vq\to0$) limit. While the dynamics of the
condensate can be characterized by some pseudo--conservation laws
governing the macroscopic phase of the order parameter (reactive
response), as well as by pair--breaking processes (dissipative
response), the system of Bogoliubov quasiparticles shows purely
relaxational behavior in the long wavelength limit. For a
quantitative study of the latter behavior we consider
impurity--limited transport, believed to dominate at low
temperatures. Collisions of the BQP with impurities are treated
within the t--matrix approximation \cite{BANDZ81,PANDP86,HVANDW86,BVANDZ06}, which is limited here to s--wave scattering. The
description thus allows one to treat the cases of weak scattering
(Born limit, $\delta_0\to 0$) and strong scattering (unitary limit
$\delta_0\to \pi/2$) on the same footing. The theory is applied to
an analysis of the electronic Raman response and the attenuation of
ultrasound in $d$--wave superconductors. Our formulation is general
enough to include the aspects of {\it universal transport}, which
was discussed previously in the literature in context with
electronic conductivity \cite{PL93}, diffusive thermal conductivity
\cite{GYSR96}, ultrasound attenuation \cite{WSS01} and electronic
Raman response \cite{WANDC98}. \\

The paper is organized as follows: After a discussion of the
equilibrium properties of unconventional superconductors in section
2, we establish a general response theory in section 3 which is
based on the classification of external perturbation potentials
through vertex functions $a_\vp$, which correspond to the specific
experiment, testing the response. Section 4 then deals with the
derivation of a two--fluid description of the response in the long
wavelength limit ($\vq\to0$) at arbitrary quasiclassical frequencies
$\omega$. The effects of the long--range Coulomb interaction are
explicitly taken into account in the derivation of the response
functions. Results for the condensate response, which are known from
the literature, are briefly rederived for completeness at the end of
this section. Section 5 is devoted to the response and relaxation
properties of the system of thermal excitations of the
superconductor, the Bogoliubov quasiparticles (BQP). It is shown
that the dynamics of the macroscopic density fluctuations of the BQP
system is entirely relaxational and describable by a set of
macroscopic relaxation times, which depend on the vertex function
$a_\vp$. This concept allows for the derivation of quite general
equations, which relate the response functions of the BQP system to
the corresponding transport parameter of given vertex function
$a_p$. The influence of the long range Coulomb interaction on the
qualitative form of the transport parameter is studied. In section 6
we consider the special case $a_\vp=1$, i. e. the relaxtion of the
macroscopic quasiparticle density in the absence of the Coulomb
renormalization and make contact to earlier work on intrinsic
density relaxation and the second viscosity. In Section 7 we derive
the explicit form of the impurity--limited relaxation time, which
enters the transport parameters using the t--matrix approximation
for the impurity self--energy. In sections 8 and 9 we discuss the
similarities in the temperature dependence of the transport
parameters associated with electronic Raman scattering and
ultrasound attenuation for various parameters characterizing
impurities in the weak (Born) and strong (unitary) scattering limit.
Section 10 is finally devoted to our summary and conclusion.
\section{Equilibrium properties} It is well established that the
pairing correlations in cuprate superconductors are {\it
unconventional} in the sense that the Fermi surface average of the
gap function $\Delta_\vp$ vanishes, i. e. \begin{eqnarray}
\langle\Delta_\vp\rangle_{\rm FS}\equiv 0 \end{eqnarray} As a
special form of the gap anisotropy we consider the case of B$_{1g}$
gap symmetry \cite{TANDK00}, \begin{eqnarray}
\Delta_\vp=\Delta_0(T)\cos(2\phi)
\end{eqnarray} Note that the nodal structure of such a gap function
implies, that the thermal excitations of the system, the Bogoliubov
quasiparticles (BQP), which have an excitation spectrum of the usual
form
\begin{eqnarray} E_\vp=\sqrt{\xi_\vp^2+\Delta_\vp^2} \ \ , \end{eqnarray}
can be created at arbitrary small energies $E_\vp$ (nodal
quasiparticles). This is reflected in the form of the BQP density of
states $N_{\rm S}(E_\vp)$ \begin{eqnarray} \frac{N_{\rm
S}(E_\vp)}{N_0}=\frac{2}{\pi}K\left(\frac{\Delta_0}{E_\vp}\right)
\stackrel{E_\vp\to 0}{=}\frac{E_\vp}{\Delta_0} \end{eqnarray}
varying linearly in the quasiparticle energy at low energy in the
clean limit. In Eq. (4) $K$ denotes the complete elliptic integral
of first kind and $N_0$ is the electronic density of states at the
Fermi surface for one spin projection. The statistical properties of
the excitation gas can conveniently be described by the thermal
Fermi--Dirac distribution
\begin{eqnarray} \nu_\vp=\frac{1}{\exp(E_\vp/\kB T)+1} \end{eqnarray} Note
that a chemical potential term is missing from (5), since the number
of thermal excitations is not fixed. \section{Response theory} In
order to test the response and transport properties of a
superconducting system, one has to apply external perturbation
potentials, which can be classified in the following way:
\begin{eqnarray} \delta\xi_\vk^{\rm
ext}=e\Phi-\frac{e}{c}\vv_\vk\cdot\vA+m\{\vM_\vk^{-1}\}_{ij}r_0A_i^IA_j^S
+\left[p_iV_{\vk
j}-\frac{\vp\cdot\vV_\vk}{d}\delta_{ij}\right]\delta u_{ij} =\sum_a
a_\vk\delta\xi_a \end{eqnarray} The first and second term in (6)
describe the coupling of the electronic system to the
electromagnetic scalar ($\Phi$) and vector ($\vA$) potential,
respectively. The third term represents the electronic coupling to a
typical Raman scattering process, with an incoming photon of energy
$\hbar\omega_I$, momentum $\hbar\vk_I$ and polarization along
$\vA^I$, and a scattered photon with energy $\hbar\omega_S$,
momentum $\hbar\vk_S$ and polarization along $\vA^S$ leaving the
sample. This process couples, for example, to electronic excitations
near the Fermi surface with energy transfer
$\hbar\omega=\hbar\omega_I-\hbar\omega_S$ and momentum transfer
$\hbar\vq=\hbar\vk_I-\hbar\vk_S$, and is describable by a
$\vk$--dependent so--called Raman tensor $\vgamma_\vk$, which we
have approximated in Eq. (6) by the inverse effective mass tensor
$\{\vM_\vk^{-1}\}_{\mu\nu}=\partial^2\xi_\vk/\hbar^2\partial k_\mu
\partial k_\nu$. In (6) $r_0=e^2/mc^2$ denotes the Thompson radius. The
fourth term in (6) represents the electronic coupling to the lattice
strain field $\delta u_{ij}$, which leads to a dissipative response
of the electronic stress tensor and hence to the attenuation of
ultrasound. In (6) $\vV_\vk=\partial
E_\vk/\hbar\partial\vk=(\xi_\vk/E_\vk)\vv_\vk$ is the group velocity
of the BQP and $d$ the dimension of the system. The r.h.s. of Eq.
(6) generalizes the $\vk$--space structure of the perturbation
potentials by introducing a $\vk$--dependent so--called vertex
function $a_\vk$, together with a collection of fictive potentials
$\delta\xi_a$, related to each vertex. In this spirit, the vertex
function $a_\vk$ is related to the specific experiment
(electromagnetic response, Raman response and relaxation, sound
attenuation) under consideration. In the Raman case one has
$a_\vk\equiv\gamma_\vk=\hat\ve_I\cdot\vgamma_\vk\cdot\hat\ve_S$ with
$\hat\ve_{I,S}$ the unit vectors in the direction of $\vA^{I,S}$ and
$\delta\xi_\gamma=r_0 |\vA^I||\vA^S|$. In the case of sound
attenuation one may write $a_\vk=\hat\vq\cdot\vsigma_\vk\cdot\hat\vu$
with the definitions $p_iV_{\vk j}-\vp\cdot\vV_\vk\delta_{ij}/d
=\pF\vvF(\xi_\vk/E_\vk)\sigma_{\vk ij}=(\xi_\vk/E_\vk)\hat{\sigma}_{\vk ij}$
and $\delta\xi_\sigma=|\vq||\vu|$
\cite{MSANDW87,MANDC96}. It is interesting to note,
that in quasi--2--$d$ systems the vertex functions $a_\vk$ for Raman
scattering and sound attenuation coincide in case of B$_{1g}$--
($a_\vk=\cos(2\phi)$) and B$_{2g}$-- ($a_\vk=\sin(2\phi)$) symmetry.
As we shall demonstrate explicitly in section 9,
these Raman polarizations can be shown to correspond to the
attenuation of {\it transverse} sound, i. e. $\hat\vq\perp\hat\vu$,
if $\vq$ is oriented parallel (B$_{2g}$--symmetry) to the crystal
($a$--) axis, or is tilted by 45$^o$ (B$_{1g}$--symmetry) from it. \\
In the homogeneous limit $\vq\to 0$ the response and transport properties
can be clearly separated into a condensate and a BQP contribution. In the
case of electronic Raman scattering, the condensate contributes to what is
referred to as the {\it pair--breaking Raman effect}, which has been extensively
discussed in the literature \cite{EANDH96,DANDK96}. In contrast,
the condensate does not contribute to dissipative processes like,
for example, the impurity--limited Raman effect or the attenuation
of ultrasound. \\

The total density fluctuation of the superconductor can, as usual,
be written as: \begin{eqnarray} \delta n_\vk(\vq,\omega)=\delta\left\langle\hat
c_{\vk+\vq\sigma}^\dagger\hat c_{\vk\sigma}\right\rangle(\omega)
\end{eqnarray} The presence of the perturbation potentials (6) gives
rise to a macroscopic density response
\begin{eqnarray}
\delta n_a(\vq,\omega)&=&\frac{1}{V}\sum_{\vp\sigma}a_\vp\delta
n_\vp(\vq,\omega) \equiv\left\langle a_\vp\delta n_\vp(\vq,\omega)\right\rangle \\
\left\langle \dots\right\rangle&=&\frac{1}{V}\sum_{\vp\sigma}\dots
\end{eqnarray}
In the presence of the long range Coulomb interaction
\begin{eqnarray} V_\vq=\frac{4\pi e^2}{\vq^2} \end{eqnarray} the density
fluctuations $\delta n(\vq,\omega)\equiv\delta n_1(\vq,\omega)$ give
rise to a molecular potential \begin{eqnarray}
\delta\xi_1=V_\vq\delta n_1(\vq,\omega) \end{eqnarray} which adds to
the external potentials $\delta\xi_\vk^{\rm ext}$ and leads to the
coupled response
\begin{eqnarray}
\delta
n_a(\vq,\omega)&=&\chi_{aa}^{(0)}(\vq,\omega)\delta\xi_a+\chi_{a1}^{(0)}(\vq,\omega)\delta\xi_1 \\ \delta
n_1(\vq,\omega)&=&\chi_{1a}^{(0)}(\vq,\omega)\delta\xi_a+\chi_{11}^{(0)}(\vq,\omega)\delta\xi_1 \nn
\end{eqnarray}
The system of Eqs. (12) can
easily be solved with the result \begin{eqnarray} \delta
n_a(\vq,\omega)&=&\chi_{aa}(\vq,\omega)\delta\xi_a  \nn \\
\chi_{aa}(\vq,\omega)&=&\chi_{aa}^{0}(\vq,\omega)-\frac{\chi_{a1}^{(0)2}(\vq,\omega)}{\chi_{11}^{(0)}(\vq,\omega)}
\left[1-\frac{1}{\epsilon(\vq,\omega)}\right] \\
\epsilon(\vq,\omega)&=&1-V_\vq\chi_{11}^{(0)}(\vq,\omega)\nn
\end{eqnarray} \section{Two--fluid description}
It can be shown that a two fluid description emerges close to the long wavelength limit
$\vq\to 0$, i. e. $\delta n_\vk(\vq\to 0,\omega)$ can be decomposed into a
condensate ($\delta n^{\rm P}_\vk$) and a quasiparticle ($\delta
n^{\rm Q}_\vk$) contribution (assumig at this stage, that the collisions are not pair--breaking):
\begin{eqnarray} \delta
n_\vk(\vq\to0,\omega)=\delta n^{\rm P}_\vk(\omega)+\delta n^{\rm
Q}_\vk(\omega) \end{eqnarray} The condensate contribution (pair
response) has the gauge--invariant form \cite{DANDE95}:
\begin{eqnarray} \delta n^{\rm P}_\vk(\omega)=-\lambda_\vk(\omega)\left\{
\delta\xi_\vk^{(+)}-\frac{i}{2}\hbar\omega\delta\varphi\right\}
\end{eqnarray} with $\lambda_\vk(\omega)$ the Tsuneto function in the long
wavelength limit
\begin{eqnarray}
\lambda_\vk(\omega)=\frac{4\Delta_\vk^2\theta_\vk}{4E_\vk^2-\omega^2}
\ \ ; \ \ \theta_\vk=\frac{1}{2E_\vk}\tanh\frac{E_\vk}{2\kB T}
\end{eqnarray}
and $\delta\xi_\vk^{(+)}$ denotes the part of (6) with $a_\vk=a_{-\vk}$.
In Eq. (15) $\delta\varphi$ denotes the
nonequilibrium phase change of the order parameter, the time
derivative of which is connected with the external perturbation
potentials of even parity through the Hamilton--Jacobi (or
generalized Josephson) relation
\begin{eqnarray}
\frac{i\hbar}{2}\omega\delta\varphi\equiv-\frac{\hbar}{2}\frac{\partial}{\partial t}\delta\varphi=
\frac{\left\langle\lambda_\vp\delta\xi_\vp^{(+)}\right\rangle}{\left\langle
\lambda_\vp\right\rangle}
=e\Phi+\sum_{a\not=1}\frac{\left\langle\lambda_\vp
a_\vp\right\rangle}{\left\langle\lambda_\vp\right\rangle}\delta\xi_a
\end{eqnarray}
where the short--hand notation (9) for the momentum sums has been used.
The physical interpretation of Eq. (17) as a Josephson
relation has been emphasized by writing out explicitly the
contribution from the scalar potential $\Phi$ to the phase change.
The other terms in (17) will turn out to vanish except for the
A$_{1g}$ Raman polarization, to be discussed later. Inserting (17)
into (15) leaves us with
\begin{eqnarray} \delta n^{\rm P}_\vk(\vq\to0,\omega)=-\lambda_\vk\left\{
\delta\xi_\vk^{(+)}-\frac{\left\langle\lambda_\vp\delta\xi_\vp^{(+)}\right
\rangle}{\left\langle\lambda_\vp\right\rangle}\right\}
=-\sum_a\lambda_\vk\left\{a_\vk-\frac{\left\langle\lambda_\vp
a_\vp\right\rangle}{\left\langle\lambda_\vp\right\rangle}\right\}\delta\xi
_a \end{eqnarray} Note that the pair response vanishes for a
constant vertex $a_\vp=$ const as a consequence of the gauge
invariance of the theory, expressed through the relation (17). The
total generalized response function $\delta n_a$ can be decomposed
into its pair (P) and quasiparticle (Q) contributions as follows
\begin{eqnarray} \delta n_a&=&\delta n_a^{\rm P}+\delta n_a^{\rm Q}
\nn \\ \delta n_a^{\rm P,Q}&=&\left\langle a_\vp\delta n_\vp^{\rm
P,Q}\right\rangle =\chi_{aa}^{{\rm P,Q}}\delta\xi_a  \\
\chi_{aa}&=&\chi_{aa}^{\rm P}+\chi_{aa}^{\rm Q}\nn \end{eqnarray}
>From (18) the renormalized pair response function $\chi_{aa}^{{\rm
P}}$ can be written in the form
\begin{eqnarray} \chi_{aa}^{{\rm
P}}&=&-\lambda_{aa}+\frac{\lambda_{a1}^2}{\lambda_{11}}  \\
\lambda_{ab}&=&\left\langle\lambda_\vp a_\vp b_\vp\right\rangle \nn
\end{eqnarray} It was shown in ref. \cite{DANDE95} that in the absence of
collisions the quantity $\Im\chi_{aa}^{{\rm P}}(\omega)$ entirely
describes the Raman response of the superconductor, the so--called
pair--breaking Raman effect. \section{Homogeneous quasiparticle
transport and relaxation} In what follows we shall therefore
concentrate on the BQP contribution (normal component in the spirit
of a two--fluid description) to the response, transport and
relaxation properties. Restricting our consideration to the case of
Raman scattering and sound attenuation, the vertex $a_\vk$ has
positive parity, i. e. $a_{-\vk}=a_\vk$. In this case, using Eq. (9),
one may define a macroscopic BQP density via
\begin{eqnarray}
\delta n^Q_a=\left\langle a_\vp \frac{\xi_\vp}{E_\vp}
\delta\nu_\vp\right\rangle \ \ ,
\end{eqnarray}
with $\delta\nu_\vp$ the deviation of the BQP distribution function from equilibrium.
Special cases include then the BQP Raman response function $\delta
n^Q_\gamma$ and the stress tensor response function
$\hat\vq\cdot\vPi^Q\cdot\hat\vu$. Here $\vPi^Q$ denotes the BQP
stress tensor and the unit vectors in the direction of propagation
($\hat\vq$) and polarization ($\hat\vu$) emerge from the standard
representation of the strain tensor $\delta u_{ij}\propto q_i u_j$
\cite{MANDC96}. It should be emphasized that the quasiparticle
stress tensor is defined as the momentum {\it current}
\begin{eqnarray}
\vPi^Q&=&\left\langle \vp:\vV_\vp
\frac{\xi_\vp}{E_\vp}h_\vp\right\rangle =
\left\langle \vp:\vv_\vp \frac{\xi_\vp^2}{E_\vp^2}h_\vp\right\rangle \nn
\\ h_\vp&=&\delta\nu_\vp+y_\vp\delta E_\vp \ \ ; \ \
y_\vk=-\frac{\partial\nu_\vk}{\partial E_\vk}\nn \\
\delta E_\vp&=&\frac{\xi_\vp}{E_\vp}\delta\xi_\vp^{\rm ext} \nn
\end{eqnarray}
and, strictly speaking, differs therefore from a generalized
density. However, we are able to show in the appendix, that whereas
the reactive response of densities and currents is indeed
qualitatively different, their dissipative response, i. e. their
transport parameters are the same. In the long wavelength limit
$\vq\to 0$, $\delta\nu_\vp$ obeys the scalar kinetic equation
\cite{WANDE78}
\begin{eqnarray}
\omega\delta\nu_\vk=i\delta I_\vk
\end{eqnarray}
where $\delta I_\vk$ represents the collision integral for the quasiparticle system.
Following ref. \cite{EANDM04}, we decompose the collision integral $\delta I_\vk$
into contributions originating from elastic (e) and inelastic (i) scattering processes:
\begin{eqnarray}
\delta I_\vk&=&\delta I_\vk^{\rm e}+\delta I_\vk^{\rm i} \nn\\
\delta I_\vk^{\rm e}&=&
-\frac{h_\vk}{\tau_\vk^{\rm e}} +\frac{y_\vk}{\tau_\vk^{\rm e}}
\frac{\xi_\vk}{E_\vk}\frac{\left\langle \frac{\xi_\vp}{E_\vp}\frac{h_\vp}{\tau_\vp^{\rm e}}\right\rangle}
{\left\langle\frac{y_\vp}{\tau_\vp^{\rm e}}\frac{\xi_\vp^2}{E_\vp^2}\right\rangle}\\
\delta I_\vk^{\rm i}&=&
-\frac{h_\vk}{\tau_\vk^{\rm i}} +\frac{y_\vk}{\tau_\vk^{\rm i}}
\frac{\xi_\vk}{E_\vk}\frac{\left\langle \frac{\xi_\vp}{E_\vp}\frac{h_\vp}{\tau_\vp^{\rm i}}\right\rangle}
{\left\langle\frac{y_\vp}{\tau_\vp^{\rm i}}\right\rangle} \nn
\end{eqnarray}
In Eq. (23) we have used approximate forms for the collision integrals,
applicable for distribution functions of positive parity $\delta\nu_{-\vk}=\delta\nu_\vk$,
which guarantee the Bogoliubov quasiparticle number conservation for elastic
scattering and allows for describing the fact that the number of Bogoliubov quasiparticles
is not conserved in context with inelastic scattering processes.
In (23) $\tau_\vp^{\rm e,i}$ denote the impurity--limited and the inelastic quasiparticle relaxation times,
respectively, of the superconductor, the first of which will be specified in more detail in section 7.
Before we perform the Coulomb renormalization, dictated by Eq. (13), it is instructive to study
the relevant response functions $\chi_{aa}^{Q(0)}$ for purely elastic scattering
\begin{eqnarray}
\chi_{aa}^{Q(0)}(\omega)_{\rm elastic}&=&-\Xi_{aa}^{\rm e}(\omega)+\frac{\Xi_{a1}^{{\rm e}2}(\omega)}{\Xi_{11}^{\rm e}(\omega)} \\
\Xi_{ab}^{\rm e}(\omega)&=&\left\langle a_\vp b_\vp\frac{\xi_\vp^2}{E_\vp^2}\frac{y_\vp}{1-i\omega\tau_\vp^{\rm e}}\right\rangle \nn
\end{eqnarray}
and purely inelastic scattering
\begin{eqnarray}
\chi_{aa}^{Q(0)}(\omega)_{\rm inelastic}&=&-\Xi_{aa}^{\rm i}(\omega)+\frac{\Xi_{a1}^{{\rm i}2}(\omega)}{\Xi_{11}^{\rm i}(\omega)}
\cdot\frac{-i\omega\tau_Q(\omega)}{1-i\omega\tau_Q(\omega)} \nn \\
\Xi_{ab}^{\rm i}(\omega)&=&\left\langle a_\vp b_\vp\frac{\xi_\vp^2}{E_\vp^2}\frac{y_\vp}{1-i\omega\tau_\vp^{\rm i}}\right\rangle \\
\tau_Q(\omega)&=&\frac{\left\langle\frac{\xi_\vp^2}{E_\vp^2}\frac{y_\vp}{1
-i\omega\tau_\vp^{\rm i}}\right\rangle}
{\left\langle\frac{\Delta_\vp^2}{E_\vp^2}\frac{y_\vp}{\tau_\vp^{\rm i}}\right\rangle} \nn
\end{eqnarray}
It is important to note that Eq. (24) expresses the number conservation law
$\chi_{11}^{Q(0)}(\omega)_{\rm elastic}=0$ for elastic scattering processes in the long wavelength limit.
For inelastic scattering, however, as represented by Eq. (25), there occurs
the well--known phenomenon of intrinsic quasiparticle relaxation \cite{WANDE78,SANDJBOOK}, described
by the lifetime $\tau_Q(\omega)$, which is finite below $\Tc$ as a consequence of the
nonconservation of the BQP number density, explicitly built into the
inelastic part of the collision integral (23). It is seen to diverge in the limit $\Delta_\vp\to 0$
since the number of quasiparticles is conserved in these processes in the normal state.
Therefore Eq. (25) is reminiscent of a viscoelastic description of
the generalized response, which interpolates between the hydrodynamic
($\omega\tau_Q\to0$) and the collisionless ($\omega\tau_Q\to\infty$)
limit.
If both elastic and inelastic scattering processes occur simultaneously, the situation
becomes more complicated, since one has to solve the integral equation
\begin{eqnarray}
\delta\nu_\vk^{(+)}&=&-\frac{y_\vk\delta E_\vk}{1-i\omega\tau_\vk^*}
+\frac{y_\vk\frac{\xi_\vk}{E_\vk}\tau_\vk^*}{1-i\omega\tau_\vk^*}
\left\{\frac{1}{\tau_\vk^{\rm e}}\frac{\left\langle
\frac{\xi_\vp}{E_\vp}\frac{h_\vp^{(+)}}{\tau_\vp^{\rm e}}\right\rangle}
{\left\langle \frac{\xi_\vp^2}{E_\vp^2}\frac{y_\vp}{\tau_\vp^{\rm e}}\right\rangle}
+\frac{1}{\tau_\vk^{\rm i}}\frac{\left\langle
\frac{\xi_\vp}{E_\vp}\frac{h_\vp^{(+)}}{\tau_\vp^{\rm i}}\right\rangle}
{\left\langle\frac{y_\vp}{\tau_\vp^{\rm i}}\right\rangle}\right\}
\end{eqnarray}
Here the index $(+)$ denotes the positive parity of the distribution
functions $\delta\nu_\vk$ and $h_\vk$ with respect to the operation $\vk\to-\vk$.
>From Eq. (26) one immediately observes, that the relaxation rates
\begin{eqnarray}
\Gamma_\vk^{\rm e,i}&=&\frac{1}{\tau_\vk^{\rm e,i}} \nn
\end{eqnarray}
do not simply add up
\begin{eqnarray}
\Gamma_\vk^*&=&\Gamma_\vk^{\rm e}+\Gamma_\vk^{\rm i}
\end{eqnarray}
to result in an effective relaxation time
\begin{eqnarray}
\tau_\vk^*&=&\frac{1}{\Gamma_\vk^*}=\frac{1}{\Gamma_\vk^{\rm e}+\Gamma_\vk^{\rm i}}
\end{eqnarray}
but there appear mixing terms originating from the collision operator in (26).
It should be noted that Eq. (26) is a straightforward generalization of the result
(3) of ref. \cite{EANDM04} to the superconducting case. Since in what follows,
we are only interested in the homogeneous limit $\vq\to 0$ of the quasiparticle
response, we may follow the argumentation of ref. \cite{EANDM04} in solving Eq. (26)
to get the final result for the full response function
$\chi_{aa}^{Q}$ (c. f. Eq. (13)) after the Coulomb renormalization:
\begin{eqnarray}
\chi_{aa}^{Q*}(\omega)&=&-\Xi_{aa}^*(\omega)+\frac{\Xi_{a1}^{*2}(\omega)}{\Xi_{11}^*
(\omega)} +O\left(\frac{1}{\epsilon}\right) \\
\Xi_{ab}^*(\omega)&=&\left\langle a_\vp b_\vp \frac{\xi_\vp^2}{E_\vp^2}\frac{y_\vp}{1-i\omega\tau_\vp^*}\right\rangle \nn
\end{eqnarray}
Note that the terms $\propto\epsilon^{-1}$ which describe the complicated mixing of elastic and inelastic
contributions (c. f. ref. \cite{EANDM04}), can be neglected in the long wavelength limit $\vq\to 0, \epsilon\to\infty$.
Therefore one may state that in the limit $\vq\to 0$ the result for the response function $\chi_{aa}^{Q}$,
which includes the effects of the long--range Coulomb interaction, has a form characteristic of a
quasiparticle number conservation law for the BQP, with $\tau_\vk^*$ entering as the effective relaxation time,
and the phenomenon of intrinsic quasiparticle relaxation becomes more or less irrelevant for charged systems
except for special experimantal situations described in chapter 5.3 of ref. \cite{SANDJBOOK}. \\

Having established the response functions of the quasiparticle
system, we would next like to clarify an important physical
consequence of the relaxation equation (22). It turns out that from
(22) one may derive a set of homogeneous relaxation equations for
the BQP densities $\delta n_a^Q$ by multiplying (22) with
$a_\vp(\xi_\vp/E_\vp)$ and summing on momentum $\vp$ and spin
$\sigma$. As a result we find that the dynamics of the BQP system is
entirely governed by relaxation processes on time scales set forth
by vertex--dependent BQP relaxation times $\tau^{Q}_{aa}$.
Assuming the fictive potentials $\delta\xi_a$ to vary as
$\propto\exp(i\vq\cdot\vr-i\omega t)$, these relaxation processes
obey a set of general equations for each vertex $a_\vk$ \cite{DALK}
\begin{eqnarray}
\omega\delta n^Q_a(\omega)=-\frac{i}{\tau^Q_{aa}(\omega)}\left[\delta
n^Q_a(\omega)-\delta n^{Q\ {\rm loc}}_a\right] \ \ \ ,
\end{eqnarray}
which describes the relaxation of the BQP density back to its local
equilibrium value \begin{eqnarray} \delta n_a^{Q\ {\rm
loc}}&=&\chi_{aa}^{Q\ {\rm loc}}\delta\xi_a \\ \chi_{aa}^{Q{\rm
loc}}&=&\chi_{ab}^{Q}(\omega=0)=-\left\langle\left(a_\vp^2-\bar{a}^2\right
) \frac{\xi_\vp^2}{E_\vp^2}y_\vp\right\rangle \nn\\
\bar{a}&=&\frac{\left\langle a_\vp \frac{\xi_\vp^2}{E_\vp^2}y_\vp
\right\rangle}
{\left\langle\frac{\xi_\vp^2}{E_\vp^2}y_\vp\right\rangle}
\end{eqnarray} From Eq. (30) we immediately get \begin{eqnarray}
\chi_{aa}^{Q}(\omega)=\frac{\chi_{aa}^{Q}(0)}{1-i\omega\tau_{aa}^{Q}(\omega)}
\end{eqnarray}
The effective quasiparticle relaxation times $\tau_{aa}^{Q}(\omega)$
are obtained as \cite{DALK} \begin{eqnarray}
\tau_{aa}^{Q}(\omega)=\frac{1}{i\omega}\left[1-\frac{\chi_{aa}^{Q}(0)}{\chi_{aa}^{Q}(\omega)}\right]
\end{eqnarray} With these results, the response function $\chi_{aa}^{Q}(\omega)$ can be decomposed
into its real and imaginary parts
as follows:
\begin{eqnarray}
\chi_{aa}^{Q}(\omega)&=&\chi_{aa}^{Q\prime}(\omega)-i\omega T_{aa}^{Q}(\omega)\nn\\
\chi_{aa}^{Q\prime}(\omega)&=& \frac{\chi_{aa}^{Q}(0)[1+\omega\tau_{aa}^{Q\prime\prime}]}
{[1+\omega\tau_{aa}^{Q\prime\prime}]^2+[\omega\tau_{aa}^{Q\prime}]^2}\\
T_{aa}^{Q}(\omega)&=&-\frac{\chi_{aa}^{Q\prime\prime}(\omega)}{\omega}=
-\frac{\chi_{aa}^{Q}(0)\tau_{aa}^{Q\prime}}
{[1+\omega\tau_{aa}^{Q\prime\prime}]^2+[\omega\tau_{aa}^{Q\prime}]^2}\nn
\end{eqnarray}
where the prime ($\prime$) and the double--prime ($\prime\prime$) refer to the real and imaginary part,
respectively. Eqs. (33, 35) describe the general connection between response and
transport of the BQP system in the homogeneous limit for a given
vertex $a_\vk$. Note that the quantity $T_{aa}^{Q}$ can be
interpreted as the generalized quasiparticle transport parameter of
the superconductor, since one may write
\begin{eqnarray} \delta
n_a^Q=\underbrace{\chi_{aa}^{Q\prime}(\omega)\delta\xi_a}_{\rm
response}+\underbrace{T_{aa}^Q(\omega)f_a}_{\rm transport}
\end{eqnarray} in which $\delta\xi_a$ and $f_a=-i\omega\delta\xi_a$
play the role of fictive potentials and forces, respectively. In the
hydrodynamic limit ($\omega\to 0$) we obtain the following result
for the effective quasiparticle relaxation times $\tau_{aa}^{Q}$
\cite{DALK}:
\begin{eqnarray}
\lim_{\omega\to 0}\tau_{aa}^{Q}(\omega)=
\frac{\left\langle\left[a_\vp-\bar{a}\right]^2\frac{\xi_\vp^2}{E_\vp^2}y_\vp
\tau_\vp^*\right\rangle}
{\left\langle\left[a_\vp^2-\bar{a}^2\right]\frac{\xi_\vp^2}{E_\vp^2}y_\vp\right\rangle}
\ ; \ \bar{a}=\frac{\left\langle a_\vp
\frac{\xi_\vp^2}{E_\vp^2}y_\vp\right\rangle} {\left\langle
\frac{\xi_\vp^2}{E_\vp^2}y_\vp\right\rangle}
\end{eqnarray} and the transport parameter $T_{aa}^{Q}$ \cite{DALK}:
\begin{eqnarray}
\lim_{\omega\to 0}T_{aa}^{Q}(\omega)=
\left\langle\left(a_\vp-\bar{a}\right)^2\frac{\xi_\vp^2}{E_\vp^2}y_\vp\tau_\vp^*\right\rangle
=\left\langle\left(a_\vp-\bar{a}\right)^2\frac{\xi_\vp^2}{E_\vp^2}
\frac{y_\vp}{\Gamma_\vp^{\rm e}+\Gamma_\vp^{\rm i}}\right\rangle
\end{eqnarray}
Note that the effects of the relaxation time $\tau_Q$ (c. f. Eq. (25)),
originating from the quasiparticle nonconservation in the inelastic scattering channel,
have completely disappeared from the result for $T_{aa}^{Q}(\omega=0)$ in the long
wavelength limit as a consequence of the long--range Coulomb interaction.
The physical consequences of intrinsic quasiparticle relaxation
can be best studied for neutral pair--correlated Fermi systems, to which we would
like to devote the following section.
\section{Intrinsic quasiparticle relaxation and second viscosity}
Although physically relevant only for neutral and not for charged systems, it is
interesting to investigate the response functions $\chi_{aa}^{Q(0)}$
and the transport parameters $T_{aa}^{Q(0)}$ in the absence of the
long--range Coulomb interaction, i. e. before the renormalization
manifested through Eqs. (12). Restricting the considerations of this section
to the clean case (i. e. purely inelastic scattering), the response function is then given by
Eq. (25) and reads in the limit $\omega\to 0$ (i. e. in local
equilibrium):
\begin{eqnarray} \chi_{aa}^{Q(0){\rm
loc}}=\chi_{ab}^{Q(0)}(\omega=0)=-\left\langle a_\vp^2
\frac{\xi_\vp^2}{E_\vp^2}y_\vp\right\rangle
\end{eqnarray}
The BQP relaxation equation is of the form (30)
\begin{eqnarray}
\omega\delta n^Q_a=-\frac{i}{\tau^{Q(0)}_{aa}}\left[\delta
n^Q_a-\delta n^{Q\ {\rm loc}}_a\right] \ \ \ ,
\end{eqnarray}
but with a different effective relaxation time
\begin{eqnarray}
\tau_{aa}^{Q(0)}(\omega)=\frac{1}{i\omega}\left[1-\frac{\chi_{aa}^{Q(0)}(0
)}{\chi_{aa}^{Q(0)}(\omega)}\right]
\end{eqnarray}
entering the representation
\begin{eqnarray}
\chi_{aa}^{Q(0)}(\omega)=\frac{\chi_{aa}^{Q(0)}(0)}{1-i\omega\tau_{aa}^{Q(
0)}(\omega)}
\end{eqnarray}
Eq. (40) represents a straightforward generalization of the problem of intrinsic BQP density relaxation
occurring in neutral Fermi superfluids \cite{WANDE78}, which can be
obtained from (40) in the special case $a_\vp\equiv 1$. The decay of
the BQP density $\delta n^Q_{1}$ occurs then as a consequence of the
fact that the number of Bogoliubov quasiparticles is not a conserved
quantity in inelastic scattering processes. Since the quasiparticle number is conserved in the normal
state, the BQP density relaxation time $\tau^Q_{11}$ must diverge in
the limit as $T\to T_{\rm c}^-$\cite{WANDE78}. In the hydrodynamic
limit ($\omega\to 0$) we obtain results for the effective
quasiparticle relaxation times $\tau_{aa}^{Q(0)}$
\begin{eqnarray}
\lim_{\omega\to 0}\tau_{aa}^{Q(0)}(\omega)= \frac{\left\langle
a_\vp^2\frac{\xi_\vp^2}{E_\vp^2}y_\vp\tau_\vp^{\rm i}\right\rangle}
{\left\langle a_\vp^2\frac{\xi_\vp^2}{E_\vp^2}y_\vp\right\rangle}
+\frac{\left\langle\frac{\xi_\vp^2}{E_\vp^2}a_\vp
y_\vp\right\rangle^2} {\left\langle a_\vp^2
\frac{\xi_\vp^2}{E_\vp^2}y_\vp\right\rangle
\left\langle\frac{\Delta_\vp^2}{E_\vp^2}\frac{y_\vp}{\tau_\vp^{\rm i}}\right\rangle}
\end{eqnarray}
and the transport parameter
$T_{aa}^{Q(0)}$:
\begin{eqnarray}
\lim_{\omega\to 0}T_{aa}^{Q(0)}(\omega)= \left\langle
a_\vp^2\frac{\xi_\vp^2}{E_\vp^2}y_\vp\tau_\vp^{\rm i}\right\rangle
+\frac{\left\langle a_\vp
\frac{\xi_\vp^2}{E_\vp^2}y_\vp\right\rangle^2}
{\left\langle\frac{\Delta_\vp^2}{E_\vp^2}\frac{y_\vp}{\tau_\vp^{\rm i}}\right\rangle}
\end{eqnarray}
which are quantitatively entirely different from the corresponding Eqs. (37) and (38).
The transport coefficient $T_{11}^{Q(0)}\equiv\zeta_3$ turns out to be nothing but
the second viscosity $\zeta_3$, which was evaluated for neutral
Fermi superfluids by W\"{o}lfle and Einzel \cite{WANDE78}:
\begin{eqnarray}
\lim_{\omega\to 0}T_{11}^{Q(0)}(\omega)&\equiv&\zeta_3=
\left\langle\frac{\xi_\vp^2}{E_\vp^2}y_\vp\tau_\vp^{\rm i}\right\rangle +\frac{\left\langle\frac{\xi_\vp^2}{E_\vp^2}
y_\vp\right\rangle^2} {\left\langle\frac{\Delta_\vp^2}{E_\vp^2}\frac{y_\vp}{\tau_\vp^{\rm i}}\right\rangle}
\end{eqnarray}
It diverges in the limit $\Delta_\vp\to 0$ as a consequence of quasiparticle number
conservation in the normal state. When applied to superfluid $^3$He--B, the inelastic
relaxation time $\tau_\vk^{\rm i}$ can be taken from the appendix 1 of ref. \cite{EANDP97}.
It should furthermore be noted, that Eq. (26) may serve and has already been used as a starting
point for a calculation of several relevant transport parameters of dirty Fermi superfluids
like $^3$He in aerogel \cite{EANDP04,EANDP05}.
\section{The resonant impurity scattering model}
The aim of this section
is to determine the relaxation time $\tau_\vp^{\rm e}$ of the superconductor in the
limit of low temperatures where the impurities play the major role
for quasiparticle scattering processes, i. e. $\tau_\vk^*\to\tau_\vk^{\rm e}$.
In that case, the relevant scattering parameters are the impurity concentration $n_i$ and the
scattering phase shift $\delta_0$ (which we would like to restrict
to the case of s--wave scattering for simplicity), giving rise to a normal
state scattering rate
\begin{eqnarray}
\frac{1}{\tau_{\rm N}}=\frac{2n_i}{\pi\hbar N_0}\sin^2\delta_0
\end{eqnarray}
In the presence of impurities, the BQP energy gets renormalized through the
impurity self--energy $\Sigma_e$ via \begin{eqnarray}
\tilde{E}_\vp=E_\vp+\Sigma_e(\tilde{E_\vp}) \end{eqnarray} with
$\Sigma_e$ evaluated within the t--matrix approximation \cite{PANDP86}
in its self--consistent version \cite{HVANDW86}:
\begin{eqnarray}
\Sigma_e(\tilde{E}_\vp)=\frac{i\hbar}{2\tau_{\rm N}}\ \
\frac{D(\tilde{E}_\vp)}{\cos^2\delta_0+\sin^2\delta_0D^2(\tilde{E}_\vp)}
\end{eqnarray}
Here, the complex function \begin{eqnarray}
D(\tilde{E}_\vp)=\left\langle\frac{\tilde{E}_\vp}{[\tilde{E}_\vp^2-\Delta_
\vp^2]^{1/2}}\right\rangle_{\rm FS} \end{eqnarray} extends the
density of states of the superconductor $N_{\rm S}(E_\vp)/N_0=\Re
D(\tilde{E}_\vp)$ to include impurity effects. From the impurity
self energy we obtain the elastic scattering rate \cite{AANDP88}
\begin{eqnarray}
\frac{1}{\tau_\vp^{\rm e}}&=&\frac{2}{\hbar}\Im\Sigma_e(\tilde{E}_\vp)= \frac{\Re
D(\tilde{E}_\vp)}{\tau_{\rm N}}\frac{\cos^2\delta_0+\sin^2\delta_0|D(\tilde{E}_\vp)|^2}
{|\cos^2\delta_0+\sin^2\delta_0D^2(\tilde{E}_\vp)|^2}
\end{eqnarray}
The explicit form for $1/\tau_\vp^{\rm e}$ becomes particularly
simple at high energies and in the limit $E_\vp\to 0$, where the
full self--consistent treatment of the renormalization
$\tilde{E}_\vk$ is necessary
($\Sigma_0^{\prime\prime}\equiv\Im\Sigma_e(0)$) \cite{HWSEP89,PL93}:
\begin{eqnarray}
\frac{\xi_\vp}{E_\vp}\tau_\vp^{\rm e}=
\begin{cases}
\Theta(E_\vp-\Delta_\vp) \frac{\tau_{\rm N}}{\Re D(E_\vp)} \left[\cos^2\delta_0+\sin^2\delta_0|D(E_\vp)|^2\right] \ ; \ E_\vp>\Sigma_0^{\prime\prime}\cr \ \ \ \ \ \ \ \ \ \ \ \ \ \ \ \ \ \ \ \ \ \ \ \frac{\hbar}{2} \frac{\Sigma^{\prime\prime 2}_0}{\left[\Sigma^{\prime\prime 2}_0+\Delta_\vp^2\right]^{3/2}} \ \ \ \ \ \ \ \ \ \ \ \ \ \ \ \ \ \ \ \ \ \ ; \ E_\vp\to 0
\end{cases}
\end{eqnarray}
\section{Impurity--limited transport} In this section we shall exploit the
general result for the transport parameter $T_{aa}^{Q}$, derived in
section 5 further and investigate its dependence on temperature and
the parameters describing the impurity scattering. Let us recall
that
\begin{eqnarray}
\lim_{\omega\to 0}T_{aa}^{Q}(\omega)&=&T_{aa}^{Q}(T)=
\frac{1}{V}\sum_{\vp\sigma}\left(a_\vp-\bar{a}\right)^2\frac{\xi_\vp^2}{E_\vp^2}y_\vp
\tau_\vp^{\rm e}
\end{eqnarray}
In what follows, we wish to restrict our considerations to the case
$\bar{a}=0$ and rewrite the transport
parameter $T^Q_{aa}$ in the following form, in which the dependence
on temperature and scattering phase shift becomes particularly
clear\cite{EANDP05}:
\begin{eqnarray}
T^Q_{aa}(T)=T^N_{aa}\left\{C_{aa}+(1-C_{aa})\left[\sin^2\delta_0Y^{(1)}_{a
a}(T)+\cos^2\delta_0Y^{(3)}_{aa}(T)\right]\right\}
\end{eqnarray}
In (53) $T^N_{aa}$ denotes the normal state limit of $T^Q_{aa}$. Eq.
(53) represents an interpolation procedure for the temperature
dependence of the transport parameter $T^Q_{aa}$, which uses the
fact, expressed in Eq. (51), that at high BQP energies the BQP
relaxation time $\tau_\vk^{\rm e}$ does not need to be evaluated
self--consistently. This gives rise to the definition of a set of
generalized Yosida functions of the form
\begin{eqnarray}
Y^{(n)}_{aa}(T)=\frac{1}{\langle a_\vp^2\rangle_{\rm FS}}
\left\langle 2\int\limits_{\Delta_\vp}^\infty dE_\vp
\frac{\sqrt{E_\vp^2-\Delta_\vp^2}}{E_\vp}
\frac{|D(E_\vp)|^{3-n}}{\Re D(E_\vp)}y_\vp a_\vp^2\right\rangle_{\rm FS}
\end{eqnarray}
\begin{figure}
  \begin{center}
  \includegraphics[width=0.75\textwidth]{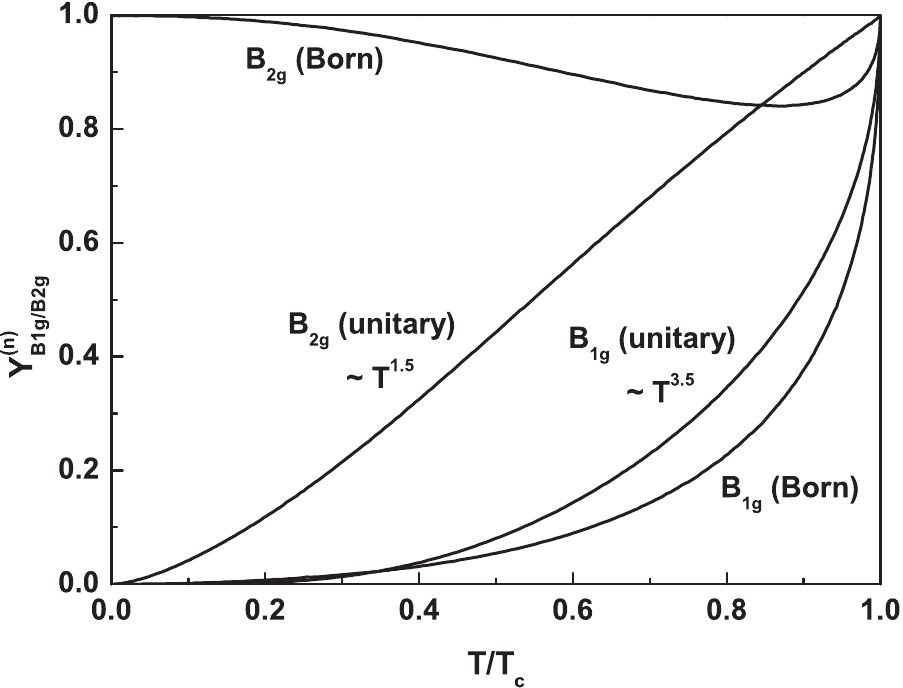}\\
  \caption{Temperature dependence of the generalized Yosida functions
  $Y^{(n)}_{aa}(T)$, characterizing the transport parameters $T_{aa}^Q$,
  for the Raman polarizations B$_{1g}$ and B$_{2g}$ in the unitary ($n=1$)
  and the Born ($n=3$) limit, as evaluated from Eq. (54) in the text.
  The stated power laws fit very well at intermediate temperatures. In the low temperature limit, even analytical
  expressions for $Y_{aa}^{(n)}(T)$ are found (c. f. Eqs. (62) and (63)).} \label{fig_01} \end{center}
\end{figure}
which describe the temperature dependence of $T^Q_{aa}$. On the
other hand, an inspection of (50) shows, that in the limit $E_\vk\to
0$ the elastic relaxation time $\tau_\vk^{\rm e}$ tends to a {\it
finite value} in the unitary limit, which reflects existence of an
impurity band originating from resonant pair--breaking processes,
described by the t--matrix. This leads to
a low--temperature offset in the temperature dependence of
$T^Q_{aa}$, which is described by the dimensionless parameter
\begin{eqnarray}
C_{aa}=\frac{\hbar}{2\tau_{\rm N}}\frac{1}{\langle
a_\vp^2\rangle_{\rm FS}}
\left\langle\frac{a_\vp^2\Sigma_0^{{\prime\prime}2}}{\left[
\Sigma_0^{{\prime\prime}2}+\Delta_\vp^2\right]^{3/2}}\right\rangle_{\rm
FS} \end{eqnarray} For a $d$--wave gap of the form (2) the integrals
over the Fermi surface can be performed in Eq. (55) in the unitary
limit with the result: \begin{eqnarray}
C_{aa}&=&\frac{2}{\pi}\frac{\hbar}{\tau_{\rm
N}\Delta_0}\frac{\Sigma_0^{\prime\prime}}{\Delta_0} \begin{cases}
K\left(\frac{i\Delta_0}{\Sigma_0^{\prime\prime}}\right)
-\frac{\Sigma_0^{\prime\prime 2}}{\Sigma_0^{\prime\prime
2}+\Delta_0^2}
E\left(\frac{i\Delta_0}{\Sigma_0^{\prime\prime}}\right) \ \ ; \ \
{\rm B}_{1g}\cr\cr
E\left(\frac{i\Delta_0}{\Sigma_0^{\prime\prime}}\right)
-K\left(\frac{i\Delta_0}{\Sigma_0^{\prime\prime}}\right) \ \ \ \ \ \ \ \ \
\ \ \  ; \ \  {\rm B}_{2g} \end{cases}\\
&\stackrel{\Sigma_0^{\prime\prime}\ll\Delta_0}{=}&\frac{2}{\pi}\frac{\hbar
}{\tau_{\rm N}\Delta_0} \begin{cases}
\left(\frac{\Sigma_0^{\prime\prime}}{\Delta_0}\right)^2
\ln\frac{4\Delta_0}{\Sigma_0^{\prime\prime}}-
\frac{\Sigma_0^{\prime\prime 2}}{\Sigma_0^{\prime\prime
2}+\Delta_0^2}\ \ ; \ \  {\rm B}_{1g}\cr\cr \ \ \
1+\frac{\Sigma^{\prime\prime 2}_0}{2\Delta_0^2}
\left[\frac{1}{2}-\ln\frac{4\Delta_0}{\Sigma^{\prime\prime}_0}\right]
 \ \  ; \ \  {\rm B}_{2g}\nn
\end{cases}
\end{eqnarray}
Here $K$ and $E$ refer to the complete elliptic integrals of first
and second kind, respectively. The description of the transport
parameter $T^Q_{aa}$ becomes {\it exact} at $T=0$ and represents a
very accurate approximation just below the transition temperature.
For intermediate temperatures, Eq. (53) represents a physically
quite transparent temperature interpolation scheme, which, however,
is meaningful only as long as the generalized Yosida functions
$Y^{(n)}_{aa}$ vanish in the low temperature limit $T\to 0$. We turn
now to an evaluation of Eq. (53) in the low temperature limit. In
the limits of unitary ($\delta_0\to\pi/2$) and Born ($\delta_0\to
0$) scattering, we obtain for not too large values of
$\Sigma^{\prime\prime}_0/\Delta_0$ (i. e. low impurity
concentrations):
\begin{eqnarray}
\lim_{T\to 0}T_{aa}^{\rm Q}&\stackrel{\delta_0\to\pi/2}{=}& \NF\langle
a_\vp^2\rangle_{\rm FS}\ \frac{2\hbar}{\pi\Delta_0}\cdot
\begin{cases}
\left(\frac{\Sigma^{\prime\prime}_0}{\Delta_0}\right)^2\ln\frac{4\Delta_0}
{\Sigma^{\prime\prime}_0} \ \   ; \ \  {\rm B}_{1g}\cr \ \ \ \ \ \ \
\ \ 1  \ \ \ \ \ \ \ \  ; \ \  {\rm B}_{2g} \end{cases} \\[1ex]
\lim_{T\to 0}T_{aa}^{\rm Q}&\stackrel{\delta_0\to 0}{=}& \NF\langle
a_\vp^2\rangle_{\rm FS}\ \frac{2\hbar}{\pi\Delta_0}\cdot
\begin{cases} \ \ \ \ 0 \ \ \ ; \ \  {\rm B}_{1g}\cr \ \ \ \ 1 \ \ ;
\ \ \ {\rm B}_{2g} \nn
\end{cases}
\end{eqnarray}
This important result shows an amazing
qualitative difference between the B$_{1g}$ and the B$_{2g}$
polarization: in the B$_{1g}$ case, the transport parameter depends
on the parameters $n_i$ and $\delta_0$, characterizing the impurity
scattering, whereas in the B$_{2g}$ case it does not. This behavior
of the zero temperature transport properties in the B$_{2g}$ case
occurs also in the case of electronic conductivity\cite{PL93}, the
electronic thermal conductivity \cite{GYSR96}, the (electron--phonon
interaction induced) sound attenuation\cite{WSS01} and has been
termed {\it universal transport}\cite{PL93}. For the case of Raman
scattering, this result has
first been derived for the unitary limit in ref. \cite{WANDC98}. \\
\begin{figure}
  \begin{center}
  \includegraphics[width=0.75\textwidth]{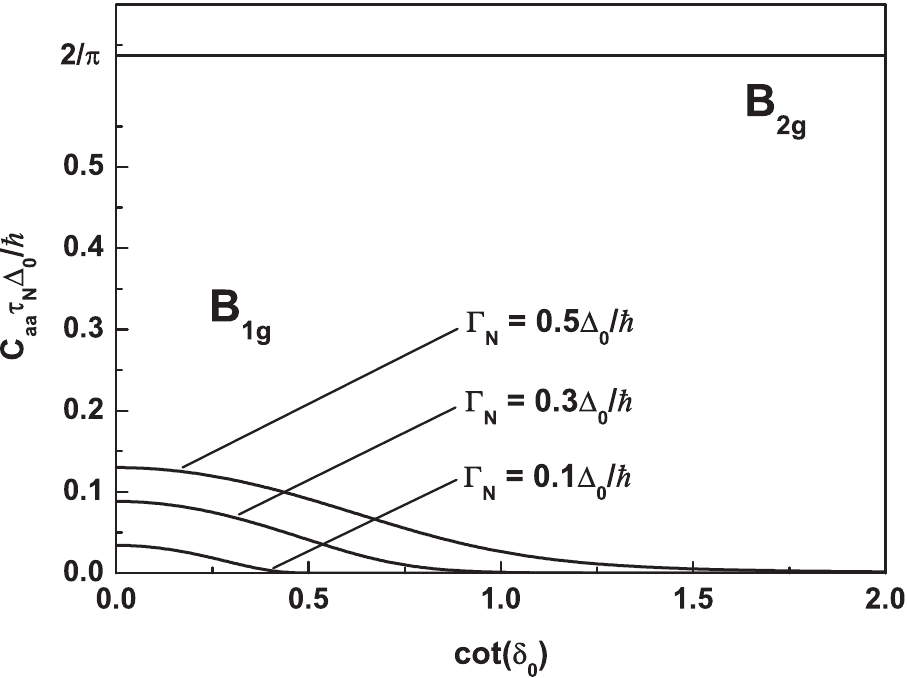}\\
  \caption{Normalized offset parameter $C_{aa}\tau_{\rm N}\propto
  T_{aa}^Q(T=0)$ giving rise to universal transport in the case of
  B$_{2g}$--symmetry ($\Gamma_N=1/\tau_N$).} \label{fig_02} \end{center}
\end{figure}
For the numerical computations it is important to see that in the
Born limit ($\delta_0\to 0$) there is a simple relation between
$\Sigma^{\prime\prime}_0$ and the normal state lifetime $\tau_{\rm
N}$, namely \begin{eqnarray} \Sigma^{\prime\prime}_0=4\Delta_0
\exp\left(-\frac{\pi\tau_{\rm N}\Delta_0}{\hbar}\right)
\end{eqnarray} In the unitary limit, on the other hand, one has to
solve the transcendental equation \begin{eqnarray}
\left(\frac{\Sigma_0^{\prime\prime}}{\Delta_0}\right)^2\ln\left(\frac{4\Delta_0}{\Sigma_0^{\prime\prime}}\right)=
\frac{\pi\hbar}{4\tau_{\rm N}\Delta_0} \end{eqnarray} in order to
relate $\Sigma^{\prime\prime}_0$ to $\tau_{\rm N}$. Defining
$\gamma=\hbar/\tau_{\rm N}\Delta_0$, the solution of Eq. (59) can be
expressed as \begin{eqnarray}
\Sigma_0^{\prime\prime}=\sqrt{\frac{\pi\gamma}{2\left|W_{-1}\left(
\frac{\pi\gamma}{32}\right)\right|}} \end{eqnarray} with $W$ the
Lambert--$W$ function, for which an expansion for small arguments
reads \cite{CGHJK96}: \begin{eqnarray}
|W_{-1}(z)|=\ln\frac{1}{z}+\ln\left(\ln\frac{1}{z}\right)\left[1+\frac{1}{
\ln\frac{1}{z}}\right]+\dots \end{eqnarray} \begin{figure}
  \begin{center}
  \includegraphics[width=0.75\textwidth]{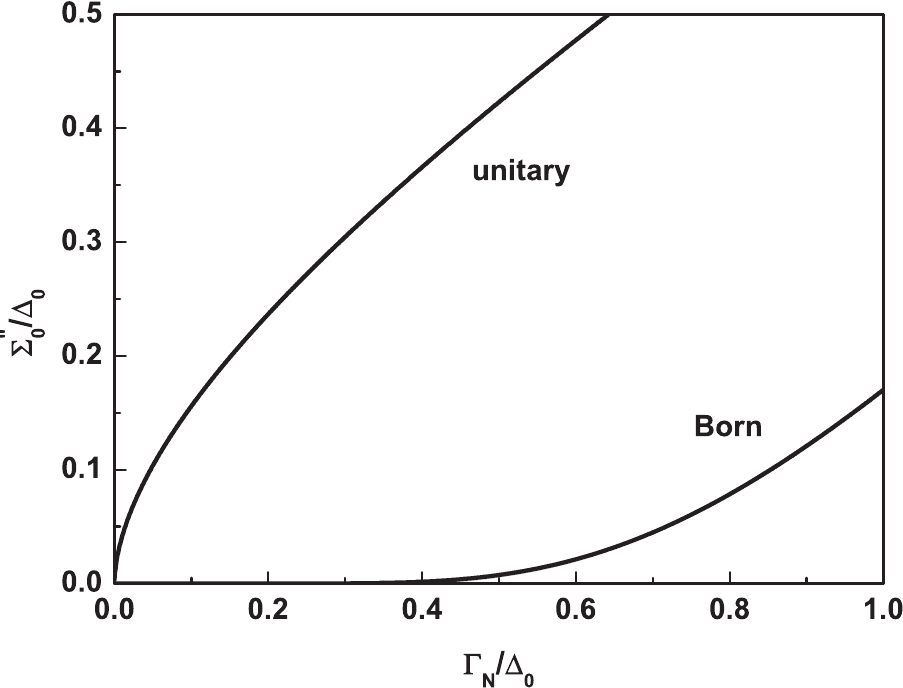}\\
  \caption{Normalized imaginary part of the impurity self energy
  $\Sigma^{\prime\prime}_0=\Im\Sigma_e(0)$ in the unitary and the Born
  limit as a function of the normalized scattering rate ${\Gamma_N=1/\tau_N}$.}
  \label{fig_03} \end{center}
\end{figure}
We consider finally the temperature dependence of $T^Q_{aa}$ in the
low temperature limit, where an analytical treatment is possible.
Using the expansion of $D(x)=x[1+i(2/\pi)\ln(x/4)]$ for
$x=E_\vk/\Delta_0\ll 1$, one obtains in the B$_{1g}$ case
\begin{eqnarray}
\lim_{T\to 0}Y^{(1)}(T)&=&4\left(\frac{\kB T}{\Delta_0}\right)^4
\left[b_{10}-b_{11}\ln\left(\frac{2\Delta_0}{\kB T}\right)
+b_{12}\ln^2\left(\frac{2\Delta_0}{\kB T}\right)\right] \\
\lim_{T\to 0}Y^{(3)}(T)&=&\frac{1}{12}\left(\frac{\pi\kB T}{\Delta_0}\right)^2 \nn
\end{eqnarray}
and in the B$_{2g}$ case
\begin{eqnarray}
\lim_{T\to 0}Y^{(1)}(T)&=&4\left(\frac{\kB T}{\Delta_0}\right)^2
\left[b_{20}-b_{21}\ln\left(\frac{2\Delta_0}{\kB T}\right)
+b_{22}\ln^2\left(\frac{2\Delta_0}{\kB T}\right)\right] \\
\lim_{T\to 0}Y^{(3)}(T)&=& 1 \nn
\end{eqnarray} Here we have defined the coefficients
\begin{eqnarray}
b_{k0}&=&a_{k0}+\frac{4a_{k2}}{\pi^2} \nn\\
b_{k1}&=&\frac{a_{k1}}{\pi^2} \\
b_{k2}&=&\frac{4a_{k0}}{\pi^2} \
\ ; \ \ k=1,2 \nn
\end{eqnarray} and
\begin{eqnarray}
a_{k\mu}=\int_0^\infty dv \frac{v^{6-2k}\ln^\mu v}{\cosh^2v}
\end{eqnarray}
In Fig. 1 we have plotted the generalized Yosida
functions $Y_{aa}^{(n)}(T)$, which characterize the
temperature-dependence of the quasiparticle transport parameters
$T_{aa}^Q$ vs. reduced temperature $T/\Tc$ [c. f. Eq. (52)].
Clearly, the theory predicts a temperature--independent result for
$T^Q_{aa}$ in the B$_{2g}$ case in the Born scattering limit. This
was actually the motivation for Pethick and Pines\cite{PANDP86} to
apply the t--matrix to the description of transport in heavy Fermion
superconductors, since the experiments showed transport parameters
vanishing at low $T$ instead of staying constant below $T_{\rm c}$.
It is quite amazing, that the $T$--dependence of
$Y^{(1)}_{aa}(T)$ is close to the power laws $T^{3.5}$ in the
B$_{1g}$-- and $T^{1.5}$ in the B$_{2g}$--case, as predicted by
Moreno and Coleman for the ultrasound attenuation\cite{MANDC96}.
In Fig. 2 we have plotted the normalized parameters $C_{aa}\tau_{\rm N}$,
which are proportional to $T_{aa}^Q(0)$, vs. $s$--wave scattering phase
shift $\cot\delta_0$ for $B_{1g}$ ($a_\vp=\cos(2\phi)$) and $B_{2g}$
($a_\vp=\sin(2\phi)$) symmetry. These parameters characterize the
low--$T$ offset in the $T$--dependence of the transport parameters
according to Eq. (52). A strong dependence of these offsets on
$\delta_0$ and $\hbar\Delta_0/\tau_{\rm N}$ is seen in the B$_{1g}$
case, whereas in the B$_{2g}$ case $T_{aa}^Q(T\to 0)$ is independent
of $\delta_0$ and $\hbar\Delta_0/\tau_{\rm N}$, at least in the
limit of low impurity concentrations $\hbar\Delta_0/\tau_{\rm N}\to 0$.
In Fig. 3 we have plotted the dependence of the quantity
$\Sigma_0^{\prime\prime}/\Delta_0$, which enters the low--$T$ offset
parameter $C_{aa}$ in Eq. (53),  as a function of the normalized
normal state scattering rate $\Gamma_{\rm N}/\Delta_0$ for both the
unitary and the Born limit of quasiparticle scattering.
\section{Raman quasiparticle transport and sound attenuation}
In this section we would like to apply the results obtained in section 8
to the transport parameters connected with the attenuation of ultrasound
and Raman scattering, and discuss strong similarities in their polarization
dependence. The vertex function $a_\vp$ reads in these two cases
for dimension $d$:
\begin{eqnarray}
a_\vp=\begin{cases}
\hat{\sigma}_\vp\equiv\pF\vvF\left[(\hat\vp\cdot\hat\vq)(\hat\vp\cdot\hat\vu)-\frac{1}{d}\hat\vq\cdot\hat\vu\right]\ \  ;
\ \ \ {\rm sound\ attenuation} \cr
\ \ \ \ \ \ \ \  \gamma_\vp\equiv m\hat{\ve}_I\cdot\vM_\vp^{-1}\cdot\hat{\ve}_S
\ \ \ \ \ \ \ \ \ \ \ \   ; \ \  {\rm Raman}
\end{cases}
\end{eqnarray}
Here the unit vectors $\hat\vq$ and $\hat\vu$ refer to the sound propagation
and polarization directions, respectively, whereas $\hat\ve_{I,S}$
denote the polarizations of the incoming and reflected photon in an electronic
Raman scattering process. Without limiting generality, we may assume
\begin{eqnarray}
\hat\vq,\hat{\ve}_I&=&\left(\begin{array}{c}
\cos \alpha \cr
\sin\alpha
\end{array} \right) \\
\hat\vu,\hat{\ve}_S&=&\left(\begin{array}{c}
\cos \beta \cr
\sin\beta
\end{array} \right) \nn
\end{eqnarray}
With this 2--$d$ representation, the stress tensor vertex and the Raman vertex assume the form
\begin{eqnarray}
\hat{\sigma}_\vp&=&\frac{\pF\vvF}{2}\left\{
\cos(\alpha+\beta)\Phi_\vp^{B_{1g}}
+\sin(\alpha+\beta)\Phi_\vp^{B_{2g}}\right\} \\
\gamma_\vp&=&\cos(\alpha-\beta)\left[
{\rm const}+\gamma_0^{A_{1g}}\Phi_\vp^{A_{1g}}\right]
+\cos(\alpha+\beta)\gamma_0^{B_{1g}}\Phi_\vp^{B_{1g}}
+\sin(\alpha+\beta)\gamma_0^{B_{2g}}\Phi_\vp^{B_{2g}}\nn
\end{eqnarray}
In (68) the quantities $\gamma_0$ are constants which depend
on the band structure and the functions $\Phi_\vp$ denote the relevant harmonics of the Fermi
surface\cite{EANDH96}. In the simplest case of a cylindrical Fermi surface one may write
\begin{eqnarray}
\Phi_\vp^{A_{1g}}=\cos(4\phi) \ \ ; \ \
\Phi_\vp^{B_{1g}}=\cos(2\phi)  \ \ ; \ \
\Phi_\vp^{B_{2g}}=\sin(2\phi)
\end{eqnarray}
Note that since by definition $\left\langle\hat{\sigma}_\vp\right\rangle_{\rm FS}=0$,
the stress tensor vertex function does not have a contribution from the
A$_{1g}$ basis function. In other respects, this appears to be the only
difference between Raman and stress tensor vertex. Note that in particular
the dependence of the transport parameters on the directions $(\hat\vq$, $\hat\vu)$ and
($\hat{\ve}_I$, $\hat{\ve}_S$) is the same. With $\hat\vx^\prime=(\hat\vx+\hat\vy)/\sqrt{2}$ and
$\hat\vy^\prime=(\hat\vx-\hat\vy)/\sqrt{2}$, this means that the attenuation
of transverse sound propagating in the $\hat\vx$-- ($\hat\vx^\prime$--) direction
corresponds to the $\hat\vx\hat\vy$-- ($\hat\vx^\prime\hat\vy^\prime$--) Raman polarization
coupling to the B$_{2g}$-- (B$_{1g}$--) basis functions. The attenuation of longitudinal sound,
on the other hand, propagating in the $\hat\vx$-- ($\hat\vx^\prime$--) direction, corresponds
to the $\hat\vx\hat\vx$-- ($\hat\vx^\prime\hat\vx^\prime$--) Raman polarization
and hence to the combinations A$_{1g}$+B$_{1g}$ (A$_{1g}$+B$_{2g}$) respectively\cite{EANDH96}.
We proceed now to write down the transport parameters associated with sound attenuation and electronic Raman
scattering. In order to be able to compare these transport phenomena, we therefore limit the
following considerations entirely to the B$_{1g}$ and B$_{2g}$ basis functions and
define generalized Yosida functions (c. f. Eq. (54))
\begin{eqnarray}
Y^{(n)}_{s}(T)&=&\frac{1}{\langle (\Phi^{s}_\vp)^2\rangle_{\rm FS}}
\left\langle  2\int\limits_{\Delta_\vp}^\infty dE_\vp
\frac{\sqrt{E_\vp^2-\Delta_\vp^2}}{E_\vp}
\frac{|D(E_\vp)|^{3-n}}{\Re D(E_\vp)}y_\vp (\Phi^{s}_\vp)^2\right\rangle_{\rm FS} \\
s&=& {\rm B}_{1g}, {\rm B}_{2g} \nn
\end{eqnarray}
and the dimensionless (offset) parameters (c. f. Eq. (55))
\begin{eqnarray}
C_{s}=\frac{\hbar}{2\tau_{\rm N}}\frac{1}{\langle
(\Phi^s_\vp)^2\rangle_{\rm FS}}
\left\langle\frac{(\Phi^s_\vp)^2\Sigma_0^{{\prime\prime}2}}{\left[
\Sigma_0^{{\prime\prime}2}+\Delta_\vp^2\right]^{3/2}}\right\rangle_{\rm FS} \ \ ; \ \
s= {\rm B}_{1g}, {\rm B}_{2g}
\end{eqnarray}
Let us start with the sound attenuation, which can be described by the wavenumber
$\alpha(\omega)$\cite{RODRIGUEZ85}:
\begin{eqnarray}
\alpha(\omega)=\frac{\omega^2}{\rho c_{\rm s}^3}\bar{\eta}
\end{eqnarray}
Here $\bar{\eta}$ is a viscous dissipation parameter, which has the following form
\begin{eqnarray}
\bar{\eta}(T)&=&\eta_{\rm N}p_s(\alpha,\beta)
\begin{cases}
C_s+(1-C_s)Y_s^{(1)}(T) \ \ ; \ \ {\rm unitary\ limit} \cr
\ \ \ \ \ \ \ \ Y_s^{(3)}(T)\ \ \ \ \ \ \ \ \ \ \ \ ; \ \ {\rm Born\ limit}
\end{cases}  \\
p_s(\alpha,\beta)&=&\cos^2(\alpha+\beta)\delta_{s,{\rm B}_{1g}}+\sin^2(\alpha+\beta)\delta_{s,{\rm B}_{2g}} \nn \\
\eta_{\rm N}&=&\frac{1}{4}n\pF\vvF\tau_{\rm N} \nn
\end{eqnarray}
In (73) $\eta_{\rm N}$ denotes the impurity--limited normal state viscosity.
The Raman case, on the other hand, is characterized by a transport parameter which is
reminiscent of the second viscosity, discussed in section 6, generalized to include the square of
the Raman vertex in the Fermi surface average:
\begin{eqnarray}
T^Q_{\gamma\gamma}(T)=\zeta_{\gamma\gamma}^{(3)}(T)
\end{eqnarray}
$\zeta_{\gamma\gamma}^{(3)}$ has the form
\begin{eqnarray}
\zeta_{\gamma\gamma}^{(3)}(T)&=&\zeta_{\gamma\gamma}^{(3){\rm N}}\left(\gamma_0^s\right)^2p_s(\alpha,\beta)
\begin{cases}
C_s+(1-C_s)Y_s^{(1)}(T) \ \ ; \ \ {\rm unitary\ limit} \cr
\ \ \ \ \ \ \ \ Y_s^{(3)}(T) \ \ \ \ \ \ \ \ \ \ \ \ ; \ \ {\rm Born\ limit}
\end{cases} \\
\zeta_{\gamma\gamma}^{(3){\rm N}}&=&\NF\tau_{\rm N}\nn
\end{eqnarray}
Hence we have demonstrated the physical similarity of the hydrodynamic transport parameters
emerging from the electronic Raman effect and the sound attenuation in $d$=2.
\section{Conclusion}
In summary, we have demonstrated, how the
response and transport properties of unconventional superconductors
are linked together in a very simple way, if the long wavelength
limit $\vq\to 0$ is taken. In this limit the contributions of the
normal component (the BQP system) and the condensate (the system of
Cooper pairs) to the response and transport simply add up and allow
for a two--fluid description which is valid at arbitrary
quasiclassical frequencies. The result for the quasiparticle
transport parameters $T_{aa}^{Q(0)}$ and $T_{aa}^{Q}$ are
qualitatively different before and after the renormalization of the
response with respect to the long--range Coulomb interaction. These
transport parameters can be represented in a form, which is valid
both in the limits of weak (Born) and strong (unitary) scattering
and which makes their dependence on temperature and the impurity
scattering parameters ($n_i, \delta_0$) particularly clear. In the
low $T$ limit, the temperature dependence can be evaluated
analytically and is found to differ from a simple power--law
behavior, predicted earlier \cite{DANDE95}. In the limit of Born
scattering, some components of the quasiparticle transport tensors
stay constant in the low $T$ limit and thus exclude the possibility
of weak scattering as a model for impurity limited transport for
example in the cuprate superconductors. In the unitary limit, on the
other hand, we find offsets in the transport parameters, which, in
certain cases, do not depend on the impurity scattering parameters.
This so--called universal transport occurs in both the
impurity--limited electronic Raman effect and the attenuation of
ultrasound. We could finally demonstrate, that for quasi--2--$d$
systems the Raman scattering intensity is closely related to the
transport of momentum (stress tensor) in the BQP system.
\section*{Appendix: Response and relaxation of densities and currents}
In
this appendix we would like to demonstrate that although one has to
carefully distinguish the response of quasiparticle densities and
currents (their reactive response is indeed different) their
dissipative response, characterized by their transport parameters,
is the same. This justifies the line of arguments presented in
section 5 of this paper. We now give a more general derivation of
response functions and transport parameters for a system of
Bogoliubov quasiparticles (BQP) in $d$--wave superconductors. The
dynamics of such a BQP system is governed by a scalar kinetic
equation for the distribution function $\delta\nu_\vk(\vq,\omega)$
(c. f. Eq. (21) in the text): \begin{eqnarray}
\omega\delta\nu_\vk-\vq\cdot\vV_\vk h_\vk&=&i\delta I_\vk
\end{eqnarray} Here $\vV_\vk=\partial E_\vk/\hbar\partial
\vk=(\xi_\vk/E_\vk)\vv_\vk$ is the BQP group velocity and the
distribution function
\begin{eqnarray}
h_\vk=\delta\nu_\vk+y_\vk\delta E_\vk
\end{eqnarray}
describes the deviation from local equilibrium. In (76) $\delta I_\vk$ is the
collision integral, which is assumed to be limited to the case of purely elastic scattering
and to have therefore the following (quasiparticle number conserving) form
\begin{eqnarray}
\delta I_\vk &=&\delta I_\vk^{(+)}+\delta
I_\vk^{(-)}\nn\\ \delta
I_\vk^{(+)}&=&-\frac{h_\vk^{(+)}}{\tau_\vk^{\rm e}}
+\frac{\xi_\vk}{E_\vk}\frac{y_\vk}{\tau_\vk^{\rm e}}
\frac{\left\langle\frac{\xi_\vp}{E_\vp}\frac{h_\vp^{(+)}}{\tau_\vp^{\rm e}}\right\rangle}
{\left\langle\frac{\xi_\vp^2}{E_\vp^2}\frac{y_\vp}{\tau_\vp^{\rm e}}\right\rangle}\\
\delta I_\vk^{(-)}&=&-\frac{h_\vk^{(-)}}{\tau_\vk^{\rm e}} \ \ ; \ \
\left\langle\dots\right\rangle=\frac{1}{V}\sum_{\vp\sigma}\dots\nn
\end{eqnarray}
Here the index $\pm$ denotes the parity of the distribution
function $h_\vk$ with respect to the operation $\vk\to-\vk$. The
first part (+) of the collision integral describes the conservation of
the BQP density, whereas the second part (--) describes the
relaxation of the BQP current, here, for simplicity, treated within
a simple relaxation time approximation. At this stage it is
important to decompose the BQP energy change $\delta E_\vk$ into
even and odd contibutions w. r. t. the operation $\vk\to-\vk$:
\begin{eqnarray}
\delta E_\vk&=&a_\vk\left[\frac{\xi_\vk}{E_\vk}\delta\xi_a+\vv_\vk\cdot\delta\vzeta_a\right]
\end{eqnarray}
In (79) the quantity $\delta\vzeta_a$ denotes a generalized
vector potential. Note that in the case of the electromagnetic response
($a_\vk\equiv e$) one has $\delta\xi_a=\Phi$ and
$\delta\vzeta_a=-\vA/c$. The macroscopic BQP density is defined
through (c. f. Eq. (21) in the text):
\begin{eqnarray} \delta n_a^{Q}=\left\langle
a_\vp\frac{\xi_\vp}{E_\vp}\delta\nu_\vp\right\rangle \end{eqnarray}
Using this definition, one may derive from (76) the following
relaxation equation for $\delta n_a^{Q}$: \begin{eqnarray}
\omega\delta n_a^Q-\vq\cdot \vj_a^Q=i\left\langle
a_\vp\frac{\xi_\vp}{E_\vp}\delta I_\vp^{(+)}\right\rangle
\end{eqnarray} Here we may identify the BQP current $\vj_a^Q$ in the
form \begin{eqnarray} \vj_a^Q&=&\left\langle
a_\vp\frac{\xi_\vp}{E_\vp}\vV_\vp h_\vp^{(-)}\right\rangle
=\left\langle a_\vp\frac{\xi_\vp^2}{E_\vp^2}\vv_\vp
h_\vp^{(-)}\right\rangle
\end{eqnarray} In the case $\langle a_\vp\rangle_{\rm FS}=0$, the solution
of the kinetic equation for the BQP density $\delta n_a^Q$,
discussed in the text, reads after invoking the effects of the
long--range Coulomb interaction (c. f. Eq. (24))
\begin{eqnarray}
\delta n_a^Q&=&\chi_{aa}^Q(\omega)\delta\xi_a \ \ ; \ \
\chi_{aa}^Q(\omega)=-\left\langle a_\vp^2\frac{\xi_\vp^2}{E_\vp^2}
\frac{y_\vp}{1-i\omega\tau_\vp^{\rm e}}\right\rangle
\end{eqnarray}
We wish now to write down the corresponding solution for the
generalized current $\vj_a^Q$. From (76) one finds for the
distribution function $h_\vk^{(-)}$ \begin{eqnarray}
h_\vk^{(-)}&=&\frac{-i\omega\tau_\vk^{\rm e}}{1-i\omega\tau_\vk^{\rm
e}}y_\vk\delta E_\vk^{(-)} -\frac{i\vq\cdot\vV_\vk\tau_\vk^{\rm
e}}{1-i\omega\tau_\vk^{\rm e}}h_\vk^{(+)} \end{eqnarray} In the long
wavelength limit $\vq\to 0$ this result, inserted into (82) leads to
\begin{eqnarray} \vj_a^Q&=&\vX_{aa}^Q(\omega)\cdot\delta\vzeta_a \ \
; \ \ \vX_{aa}^Q(\omega)=\left\langle
a_\vp^2\frac{\xi_\vp^2}{E_\vp^2}\vv_\vp:\vv_\vp
\frac{-i\omega\tau_\vp^{\rm e}}{1-i\omega\tau_\vp^{\rm
e}}y_\vp\right\rangle \end{eqnarray} One immediately recognizes that
the response functions for the BQP density $\chi_{aa}^Q(\omega)$ and
the quasiparticle current $\vX_{aa}^Q(\omega)$ differ, besides the
tensor structure of the latter, in the usual way in their dependence
on the dimensionless quantity $\omega\tau_\vp^{\rm e}$. It is
instructive to decompose the density response function $\chi_{aa}^Q$
into its real and imaginary parts, respectively:
\begin{eqnarray}
\chi_{aa}^Q&=&\chi_{aa}^{Q\prime}+i\chi_{aa}^{Q\prime\prime}\nn \\
\chi_{aa}^{Q\prime}&=&-\left\langle a_\vp^2\frac{\xi_\vp^2}{E_\vp^2}
\frac{y_\vp}{1+(\omega\tau_\vp^{\rm e})^2}\right\rangle \\
\chi_{aa}^{Q\prime\prime}&=&-\omega\left\langle
a_\vp^2\frac{\xi_\vp^2}{E_\vp^2} \frac{y_\vp\tau_\vp^{\rm
e}}{1+(\omega\tau_\vp^{\rm e})^2}\right\rangle \equiv-\omega
T_{aa}^Q\nn
\end{eqnarray}
In the same way we obtain for $\vX_{aa}^Q$:
\begin{eqnarray}
\vX_{aa}^Q&=&\vX_{aa}^{Q\prime}+i\vX_{aa}^{Q\prime\prime}\nn \\
\vX_{aa}^{Q\prime}&=&\left\langle
a_\vp^2\frac{\xi_\vp^2}{E_\vp^2}\vv_\vp:\vv_\vp
\frac{y_\vp(\omega\tau_\vp^{\rm e})^2}{1+(\omega\tau_\vp^{\rm
e})^2}\right\rangle \\
\vX_{aa}^{Q\prime\prime}&=&-\omega\left\langle
a_\vp^2\frac{\xi_\vp^2}{E_\vp^2}\vv_\vp:\vv_\vp
\frac{y_\vp\tau_\vp^{\rm e}}{1+(\omega\tau_\vp^{\rm
e})^2}\right\rangle \equiv-\omega \vT_{aa}^Q\nn
\end{eqnarray}
It is interesting to formally compare the response of the
BQP density and current in the following way: The density response
can be written in terms of an effective BQP relaxation time
$\tau_{aa}^Q$ in the form (c. f. Eq. (33) in the text):
\begin{eqnarray}
\delta n_a^Q&=&-\frac{\left\langle
a_\vp^2\frac{\xi_\vp^2}{E_\vp^2}y_\vp\right\rangle}
{1-i\omega\tau_{aa}^Q}\delta\xi_a \ \ ; \ \
\tau_{aa}^Q=\frac{\left\langle
a_\vp^2\frac{\xi_\vp^2}{E_\vp^2}\frac{y_\vp\tau_\vp^{\rm e}}
{1-i\omega\tau_\vp^{\rm e}}\right\rangle} {\left\langle
a_\vp^2\frac{\xi_\vp^2}{E_\vp^2}\frac{y_\vp} {1-i\omega\tau_\vp^{\rm
e}}\right\rangle} \end{eqnarray}
whereas the current obeys a Drude--like law, involving a corresponding  effective current
relaxation time $\tau_{avav}^Q$:
\begin{eqnarray}
\vj_a^Q&=&\frac{-i\omega\tau_{avav}^Q}{1-i\omega\tau_{avav}^Q}
\left\langle
a_\vp^2\vv_\vp\frac{\xi_\vp^2}{E_\vp^2}y_\vp(\vv_\vp\cdot\delta\vzeta_a)\right\rangle\
\ ; \ \ \tau_{avav}^Q=\frac{\left\langle
(a_\vp\vv_\vp)^2\frac{\xi_\vp^2}{E_\vp^2}\frac{y_\vp\tau_\vp^{\rm
e}} {1-i\omega\tau_\vp^{\rm e}}\right\rangle} {\left\langle
(a_\vp\vv_\vp)^2\frac{\xi_\vp^2}{E_\vp^2}\frac{y_\vp}
{1-i\omega\tau_\vp^{\rm e}}\right\rangle}
\end{eqnarray}
Eqs. (88) and (89) are clearly seen to describe the formal difference between
the density and current response: whereas the effective relaxation
times for the densities and currents differ only by the use of
different vertices $a_\vp\leftrightarrow a_\vp\vv_\vp$, the response
of the current is Drude--like, whereas the response of the density
is not. Nevertheless we have now arrived at a stage where we can
analyze the connection between the density and current response
functions and the transport parameters associated with the densities
and currents. Defining generalized forces
\begin{eqnarray}
f_a&=&-i\omega \delta\xi_a \ \ ; \ \ \vf_a=-i\omega
\delta\vzeta_a \end{eqnarray} we may write \begin{eqnarray} \delta
n_a^Q&=&\chi_{aa}^{Q\prime}\delta\xi_a+T_{aa}^Q f_a \\
\vj_a^Q&=&\vX_{aa}^{Q\prime}\cdot\delta\vzeta_a+\vT_{aa}^Q\cdot\vf_a\nn
\end{eqnarray}
An inspection of Eqs. (83, 85, 88, 89)  shows that while
the reactive response of BQP densities and currents is qualitatively
different, the dissipative response of densities and currents,
represented by the transport parameters $T_{aa}^Q$ and $\vT_{aa}^Q$
is similar in structure, if on makes the replacements for the
vertices $a_\vp\to a_\vp\vv_\vp$. The transport parameters
$T_{aa}^Q$ and $\vT_{aa}^Q$ read
\begin{eqnarray}
T_{aa}^Q(\omega)&=&-\frac{\chi_{aa}^{Q\prime\prime}(\omega)}{\omega}
=\left\langle a_\vp^2\frac{\xi_\vp^2}{E_\vp^2}
\frac{y_\vp\tau_\vp^{\rm
e}}{1+(\omega\tau_\vp^{\rm e})^2}\right\rangle \\
\vT_{aa}^Q(\omega)&=&-\frac{\vX_{aa}^{Q\prime\prime}(\omega)}{\omega}
=\left\langle a_\vp^2\frac{\xi_\vp^2}{E_\vp^2}\vv_\vp:\vv_\vp
\frac{y_\vp\tau_\vp^{\rm e}}{1+(\omega\tau_\vp^{\rm
e})^2}\right\rangle \nn
\end{eqnarray}
Hence we have arrived at a unified description of transport phenomena associated with the Raman
and stress tensor response and have therefore justified the argumentation presented in section 5.
\section*{Acknowledgments}
Enlightening discussions with and helpful remarks from Rudi Hackl, Dirk Manske, Leonardo Tassini,
Johannes Waldmann, Peter W\"olfle and Fred Zawadowski are gratefully acknowledged.

\end{document}